\documentclass[aps,prl,reprint,superscriptaddress,longbibliography,nofootinbib]{revtex4-2}

\usepackage{amsmath,amssymb}
\usepackage{braket}
\usepackage{graphicx}
\usepackage{xcolor}
\usepackage{hyperref}
\hypersetup{
 colorlinks=true,
 linkcolor=blue!60!black,
 citecolor=blue!60!black,
 urlcolor=blue!60!black
}
\newcommand{\cre}{\mathrm{QRE}}
\newcommand{\en}{\mathrm{EN}}

\begin{document}
\title{Quantum Residual Exploration for Selected Configuration Interaction}
\author{Kentaro Yamamoto}
\affiliation{Quantinuum K.K., Otemachi Financial City Grand Cube 3F, 1-9-2 Otemachi, Chiyoda-ku, Tokyo, Japan}
\author{Ikko Hamamura}
\affiliation{NVIDIA Corporation, Tokyo 107-0052, Japan}
\author{Juan W. Pedersen}
\affiliation{Quantinuum K.K., Otemachi Financial City Grand Cube 3F, 1-9-2 Otemachi, Chiyoda-ku, Tokyo, Japan}
\author{Duncan Gowland}
\affiliation{Quantinuum, Terrington House, 13-15 Hills Road, Cambridge CB2 1NL, UK}
\author{Gabriel Laude}
\affiliation{Quantinuum, Terrington House, 13-15 Hills Road, Cambridge CB2 1NL, UK}
\author{Shintaro Maekawa}
\affiliation{Quantinuum K.K., Otemachi Financial City Grand Cube 3F, 1-9-2 Otemachi, Chiyoda-ku, Tokyo, Japan}
\author{David Mu\~{n}oz Ramo}
\affiliation{Quantinuum, Terrington House, 13-15 Hills Road, Cambridge CB2 1NL, UK}
\author{Elica Kyoseva}
\affiliation{NVIDIA Corporation, Santa Clara, CA 95051, USA}
\author{William Dawson}
\affiliation{RIKEN Center for Computational Science, RIKEN, Kobe, Hyogo 650-0047, Japan}
\date{November 9, 2026}

\begin{abstract}
We propose a modular hybrid quantum-classical method to improve selected configuration
interaction at a fixed final determinant budget. Quantum measurements supply complementary
determinants, accepted only when they outrank classical candidates under a
common energetic criterion. This targets multistep excitations that
perturbative discovery may miss in strongly correlated systems.
A demonstration on the Quantinuum Reimei trapped-ion quantum computer shows
that measured proposals for H$_{12}$ at a stretched geometry lower the
variational energy relative to classical-only selection at the same final
determinant budget.
\end{abstract}
\maketitle

Classical selected configuration interaction (SCI) methods have achieved
remarkable success in electronic-structure
calculations~\cite{Huron1973,Tubman2016,Holmes2016,Schriber2016ACI,
Liu2016ICI,Sharma2017,Helgaker2013,Zhang2025TrimCI}.
By retaining selected Slater determinants and diagonalizing the Hamiltonian
in the subspace they span, SCI makes calculations tractable despite the combinatorial
growth of the full configuration space. Its accuracy within fixed classical
memory and compute budgets depends on which determinants occupy that space. Effective determinant
discovery is therefore central to exploiting the available classical resources.

Classical determinant discovery commonly uses perturbatively motivated
importance estimates based on direct Hamiltonian couplings to the selected
space~\cite{Huron1973,Holmes2016}.
Strongly correlated systems can require substantial contributions from
multiple configurations~\cite{Holmes2016,Motta2017}.
The electronic Hamiltonian couples determinants through at most two
substitutions~\cite{Helgaker2013}, so important determinants may lack direct
connections to the initial selected space. Classical SCI can discover them
through repeated selection and diagonalization~\cite{Huron1973,Holmes2016},
although discovery guided by local importance estimates may be inefficient
within a finite proposal budget.

Kanno et al.~\cite{Kanno2026} proposed quantum-selected configuration
interaction (QSCI), offering an alternative route to determinant discovery
through quantum sampling. Subsequent QSCI
developments~\cite{Nakagawa2024ADAPTQSCI,Sugisaki2025HSBQSCI,Loetstedt2026,Yamamoto2026}
and related approaches, including sample-based quantum
diagonalization~\cite{RobledoMoreno2025,Barison2025,Liepuoniute2025,Shirakawa2025,Kumar2026BSSQD},
use quantum circuits to identify configurations for constructing an effective
CI subspace.
Reinholdt et al.~\cite{Reinholdt2025} found that, for the systems studied,
sampling from the ground-state distribution incurred substantial
repeated-sampling overhead or required larger CI expansions than Heat-bath configuration interaction (HCI) for
comparable energy accuracy.
This motivates us to explore
the incremental utility of quantum sampling as a complementary source of
determinants.

We introduce quantum residual exploration (QRE) for this purpose. Its
quantum proposal state is designed to explore excitation pathways that local perturbative
selection from an established classical SCI space may reach only after
further iterations. The state need not approximate the target eigenstate;
measured determinants are ranked alongside classical proposals under a common
Hamiltonian-based criterion before selection into the final CI space.
The present benchmark assesses the incremental energy improvement from
quantum proposals over classical-only selection at the same final determinant
budget, rather than convergence to chemical accuracy.

To assess whether quantum-generated proposals retain their utility under
device noise and finite sampling, we perform a hardware demonstration for an
H$_{12}$ chain on Quantinuum Reimei, a trapped-ion quantum computer.
A separate noiseless benchmark examines how direct and multistep
contributions change across bond lengths.

\begin{figure}[t]
 \centering
 \includegraphics[width=\columnwidth]{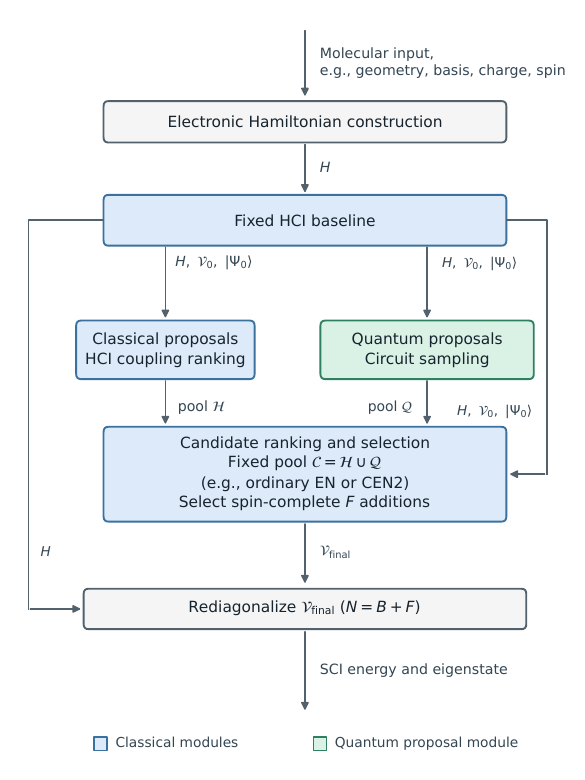}
 \caption{Modular quantum-assisted selected-CI workflow with QRE. Molecular inputs
 define the electronic Hamiltonian $H$, and a fixed HCI baseline supplies
 the classical and quantum proposal branches. The baseline supplies
 information for scoring the transport generators.
 Their fixed pools form
 $\mathcal C=\mathcal H\cup\mathcal Q$, which undergoes one ranking and
 spin-complete selection to produce $\mathcal V_{\mathrm{final}}$ at dimension
 $N=B+F$. Final variational
 diagonalization is performed to obtain the ground-state energy and eigenstate.
 }
 \label{fig:workflow}
\end{figure}

The present QRE workflow is summarized in Fig.~\ref{fig:workflow}. It starts by
constructing the electronic Hamiltonian from molecular input. A fixed HCI
baseline then feeds classical and quantum proposal branches. Their candidates
undergo a common ranking and spin-complete selection, followed by one
variational diagonalization in the final determinant space. Below, we describe the baseline, proposal generation, ranking, and final
diagonalization in the order shown in Fig.~\ref{fig:workflow}. These components
can be refined independently within QRE. Implementation details and supporting
comparisons for the workflow components below are given in Supplemental
Material, Sec.~S1.

\textit{Electronic Hamiltonian construction---} Given a molecular input such as nuclear geometry, charge, and basis set, classical
preprocessing evaluates the one- and two-body integrals defining the
second-quantized electronic Hamiltonian $H$~\cite{Helgaker2013}. The electron
number and spin sector specify the determinant space.

\textit{Fixed HCI baseline---} We use heat-bath configuration interaction
(HCI)~\cite{Holmes2016,Sharma2017} as a representative classical SCI method
with an efficient and explicit determinant-selection criterion. This provides
a reproducible baseline for assessing the incremental utility of quantum
proposals. Let $\mathcal V$ be a variational
determinant space and $C_J(\mathcal V)$ the coefficient of $\ket{D_J}$ in its
normalized CI solution. Here $\ket{D_J}$ is a Slater determinant and
$H_{IJ}=\bra{D_I}H\ket{D_J}$ a matrix element of $H$; $I$ and $J$ label
determinants. Following the HCI selection criterion of Holmes et
al.~\cite{Holmes2016}, an external determinant $I\notin\mathcal V$ has score
\begin{equation} S_I^{\mathrm{HCI}}(\mathcal V)=\max_{J\in\mathcal
V}|H_{IJ}C_J(\mathcal V)|. \label{eq:hci} \end{equation} During iterative
baseline construction, $\mathcal V$ and its CI coefficients are updated after
each selection--diagonalization step; candidates with
$S_I^{\mathrm{HCI}}(\mathcal V)>\epsilon_0$ are added with spin completion.

We construct the baseline using HCI with a coarse selection threshold,
leaving room for subsequent refinement.
The resulting space $\mathcal V_0$ defines the baseline size
$B=|\mathcal V_0|$. Its normalized CI solution is
\[
 \ket{\Psi_0}=\sum_{J\in\mathcal V_0}C_J^{(0)}\ket{D_J},
\]
with energy $E_0$.

\textit{Quantum proposals---} The design principle is to use a quantum
proposal state to discover determinants that classical HCI may struggle to
find within a finite proposal budget.
In the implementation studied here,
coherent evolution transfers amplitude from a reference state along excitation
pathways selected using classical importance estimates. A midcircuit
measurement rejects configurations that retain the reference occupation
pattern on selected frozen orbitals. The remaining gates further explore
the accepted configurations.

\begin{figure}[t]
 \centering
 \includegraphics[width=\columnwidth]{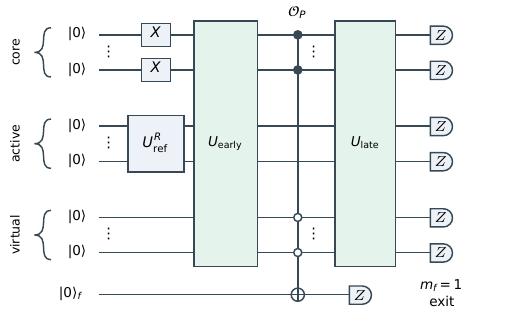}
 \caption{An example QRE proposal circuit used in this work. Ordered transport
 blocks surround a flag
 measurement that rejects a selected frozen-occupation pattern. Filled/open
 controls require occupied/empty qubits; the accepted $m_f=0$ branch retains
 coherence before late evolution. Final measurements yield candidate
 occupation strings.}
 \label{fig:projected-cre-circuit}
\end{figure}

Fig.~\ref{fig:projected-cre-circuit} shows the example circuit used in this work,
whose accepted state is
\begin{equation}
 \ket{\psi_{\cre}}=
 \frac{U_M\cdots U_{k+1}Q U_k\cdots U_1\ket{\Phi^R}}{\sqrt{p_Q}}.
 \label{eq:projected-cre}
\end{equation}
Starting from the reference $\ket{\Phi^R}$ in the target orbital
space, with determinant support $\mathcal R$, the circuit applies $M$ ordered
transport gates $U_\mu=e^{\theta_\mu G_\mu}$. Here
$G_\mu=T_\mu-T_\mu^\dagger$ is the anti-Hermitian generator associated with a
spin-adapted single or double excitation. After gate $k$, a membership
measurement retains configurations that differ from the reference occupation
pattern on at least one selected frozen orbital, defining the sector
projected by $Q$; its acceptance probability
$p_Q=\|Q U_k\cdots U_1\ket{\Phi^R}\|^2$ normalizes the accepted state.

In the implementation studied here, a reduced reference is constructed
classically from a complete-active-space CI (CASCI) ground state in a smaller
active space and embedded in the target orbital space~\cite{Yamamoto2026}.
This CASCI reference initializes the quantum circuit, while the HCI baseline
supplies the residuals used to score its transport generators.
The statevector benchmark uses the reference coefficients directly; its
circuit settings are given in End Matter.

The correlated warm start is one realization of this design principle.
A statevector comparison replaces it with a Hartree--Fock determinant while
retaining the same transport circuit.
This simpler input still yields an incremental energy improvement, although
the correlated reference gives a larger gain.

To choose the ordered gates $U_1,\ldots,U_M$, we first form eligible
single- and double-excitation generators from reference-to-baseline
Hamiltonian connections. We then assess
each generator through the external determinants it can reach directly from
$\ket{\Phi^R}$, combining their accessibility with their Epstein--Nesbet (EN) importance to
the fixed baseline. For every external determinant $\ket{D_I}$ with
$I\notin\mathcal V_0$, the baseline state defines the direct
residual and signed EN denominator~\cite{Epstein1926,Nesbet1955}.\footnote{%
In the numerical implementation, vanishing denominators are regularized:
all scores and amplitudes use $\Delta_I=-\operatorname{sgn}(H_{II}-E_0)
\max\{|H_{II}-E_0|,\delta\}$ with $\delta=10^{-12}$~Ha
and $\operatorname{sgn}(0)=+1$.}
\begin{equation}
 r_I=\sum_{J\in\mathcal V_0}H_{IJ}C_J^{(0)},\qquad
 \Delta_I=E_0-H_{II}.
 \label{eq:en}
\end{equation}
The amplitude $q_{I\mu}=\bra{D_I}G_\mu\ket{\Phi^R}$ measures direct
accessibility from the reference under generator $\mu$. We combine it with
the ordinary EN score $|r_I|^2/|\Delta_I|$ to rank generators:
\begin{equation}
 w_\mu=\sum_{I\notin\mathcal V_0\cup\mathcal R}
 |q_{I\mu}|\frac{|r_I|^2}{|\Delta_I|}.
 \label{eq:generator-score}
\end{equation}
We retain the $M$ highest-scoring generators and apply them in descending
$w_\mu$ order. This one-generator metric chooses the circuit; the sequence
can reach further configurations, and final candidate admission
uses the separate classical ranking procedure below.
At $R=1.8$~\AA, varying the common transport angle from 0.10 to 0.20~rad
changes the QRE+CEN2 energy improvement by less than 0.6~mHa.

An auxiliary flag qubit implements the membership measurement; retaining
outcome $m_f=0$ applies $Q$ while preserving coherence among accepted
configurations.
Postselection incurs an ideal acceptance overhead of $1/p_Q$; rejected
preparations count toward the attempted-shot budget. Its net benefit depends
on circuit cost and the number of useful distinct proposals retained.
Early termination, supported by platforms such as Quantinuum
Helios~\cite{Ransford2026Helios}, could skip the remaining transport gates
on rejected shots without changing the ideal accepted proposal distribution.

Quantum measurements provide candidate determinants; the classical selection
procedure requires only these candidates, irrespective of the circuit used to
produce them. This
separates the design of the quantum proposal distribution from the classical
assessment of its energetic value. These candidates are processed into
symmetry-valid, spin-complete candidate sets and combined with classical
proposals for common energetic ranking. The processing rules are given in
End Matter.

\textit{Classical proposals---} To provide classical candidates for comparison
with the quantum proposals, we generate a finite one-shot pool indexed by
$\mathcal H$. Spin-complete bundles are ranked using the HCI score evaluated
at $\mathcal V=\mathcal V_0$, where $C_J(\mathcal V_0)=C_J^{(0)}$.
The baseline threshold is not reapplied, and no intermediate diagonalization
is performed.

\textit{Candidate ranking and selection---} Ordinary EN~\cite{Huron1973}
is a natural ranking choice because $S_I^{\en}=|r_I|^2/|\Delta_I|$ measures
direct coupling to the fixed baseline. However, this score vanishes when
$r_I=0$. Such a determinant can still lower the variational energy when
selected together with other external determinants to which it couples.
This matters for QRE because its principal expected contribution
is to propose determinants reached through multistep excitation pathways.
To assess such candidates within a one-shot ranking, we define a second-order
collective EN score (CEN2) using two signed propagation updates within a
fixed candidate pool $\mathcal P$, starting from the direct baseline residual.
CEN2 is the lowest-order one-shot choice
that can score paths through two external intermediates; their contributions
may reinforce or cancel. Candidates with no contributing path within this
reach retain zero individual score, although a positive-score spin-bundle
partner may admit them. Longer paths are not explicitly included in the CEN2
score; couplings among the selected determinants are treated in the final
variational diagonalization. CEN2 remains
a modular choice rather than a requirement of QRE. Here
$\mathcal P=\mathcal C$ for QRE and
$\mathcal P=\mathcal H\cup(\mathcal R\setminus\mathcal V_0)$ for the classical-only
control.

Spin-complete bundles are ranked once by the sum of their determinant scores.
Let $\mathcal A$ be the set of $F\in\mathbb N$ admitted determinants.
Classical and quantum proposals compete for the same $F$ additions under
the common ranking criterion.

\textit{Final diagonalization---} The final selected space is
$\mathcal V_{\mathrm{final}}=\mathcal V_0\cup\mathcal A$, with
$|\mathcal V_{\mathrm{final}}|=N=B+F$. We rediagonalize this space once.
The main-text figures show variational energies. We also apply Epstein--Nesbet
second-order perturbation theory (EN-PT2)~\cite{Huron1973} to the final
HCI+CEN2 and QRE+CEN2 states in the statevector benchmark. Each correction
includes all Hamiltonian-connected external determinants.
The correction procedure is detailed in End Matter.

We use an equally spaced linear H$_{12}$ chain in STO-3G as a controlled
demonstration of when QRE proposals can add value to classical determinant
selection.
Stretching the chain increases the importance of multiple electronic
configurations and its expected multireference character~\cite{Motta2017}.
The bond-length scan therefore tests how the incremental benefit of QRE
changes as electronic correlation becomes stronger.

To assess proposal sources on common footing, we first fix the molecular and
classical CI setup. The fixed-$M_S=0$ CAS(12,12) space contains
$\binom{12}{6}^2=853{,}776$ Slater determinants; full configuration interaction
(FCI) provides the exact energy within this basis and a proposal source for
comparison, as described below. We truncate the centered CAS(8,8) ground state to retain at least 99\% of
its probability weight, complete its spin configurations, and embed it in
CAS(12,12). This reference initializes QRE and seeds the HCI baseline. At $\epsilon_0=2.3$~mHa, this baseline grows
from $B=3{,}703$ to 14,463 determinants over the scan. Each branch retains the
same baseline at a given geometry and performs one ranking and selection
followed by final diagonalization at $N=20{,}000$. The common classical HCI
proposal pool has capacity 30,000, and reduced-reference support outside the
baseline is included in every branch.

For the present demonstration, we use ten transport gates and place the
projection after the fifth gate. This midpoint placement is fixed rather
than optimized.

The statevector benchmark covers seven H--H separations from 1.2 to 1.8~\AA.
At each geometry, the normalized coefficients of the truncated reference
initialize the QRE circuit, and exact probabilities of the resulting
postselected state define proposals.
To assess ideal coverage of the external determinant space, proposal pools
are constructed from exact probabilities after excluding fixed-baseline
and reference support. This comparison does not assign a measurement-shot
cost; repeated outcomes and already-known determinants still consume shots
in practice, as examined below in the finite-shot analysis.
We compare QRE to including an equal-cardinality set of random determinants
at each geometry. Both augment the HCI pool with up to 30,000 determinants. The baseline sizes for the different geometries leave 16,297 to 5,537
additions at the fixed final dimension, making this a test of determinant discovery rather
than a convergence study for H$_{12}$. Implementation details and intermediate
benchmark data are given in Supplemental Material, Sec.~S1.

We use one HCI proposal stage to avoid additional intermediate
diagonalizations. A five-generation HCI diagnostic preserves the qualitative
advantage of QRE proposals at stretched geometries (Supplemental Material,
Sec.~S2).

As a control, we replace the QRE proposal probabilities with exact FCI
probabilities, prioritizing spin-complete groups with the largest weight in
the FCI ground state. Everything else remains unchanged.

The hardware demonstration focuses on the stretched geometry $R=1.8$~\AA\
and uses Quantinuum Reimei, a 56-qubit trapped-ion quantum
computer, together with ROQUO, a GPU
supercomputer comprising 135 NVIDIA GB200 NVL4 nodes with 540 Blackwell
GPUs. Both systems are part of RIKEN's quantum--HPC hybrid
platform~\cite{RIKENROQUO2026}. The quantum circuit follows
Fig.~\ref{fig:projected-cre-circuit} and uses the Jordan--Wigner mapping with
24 system qubits. The projector checks the occupations of the eight frozen
spin orbitals through a multi-ancilla implementation of an eight-controlled
NOT gate~\cite{Barenco1995}. Six work qubits store intermediate logical ANDs,
and one membership-flag qubit records the result, bringing the total to
31 qubits. The flag measurement retains the component outside the
embedded-reference sector before the remaining transport gates are applied.

Synthesizing a state-preparation circuit for an arbitrary classically
specified reference state is a nontrivial task. We therefore use an
evolutionary algorithm to search for a compact circuit that reaches the
target reference-state fidelity within the resource budget of a practical hardware
experiment.
This preparation objective concerns the reduced reference,
rather than the target-space FCI eigenstate. The preparation procedure is
detailed in End Matter.
Related work combines evolutionary search with large language models to
discover quantum algorithms for molecular ground states~\cite{Finger2026Discovery}.
Circuit components are optimized separately, without optimization across the
projector, and compiled into native single- and two-qubit gates. The full
circuit contains 939 two-qubit gates; compilation details are given in End Matter.
A total of 10,000 attempted shots supplies measured
candidates after flag and symmetry filtering; these are
ranked against the fixed HCI pool using the same fixed baseline and EN or CEN2
rule as the statevector branches.
Benchmark details and proposal-source controls are given in Supplemental
Material, Sec.~S1; hardware implementation and device specifications
are given in End Matter.
Software packages, analysis-environment versions, and workflow tools are
listed in End Matter.

\begin{figure}[h]
 \centering
 \includegraphics[width=\columnwidth]{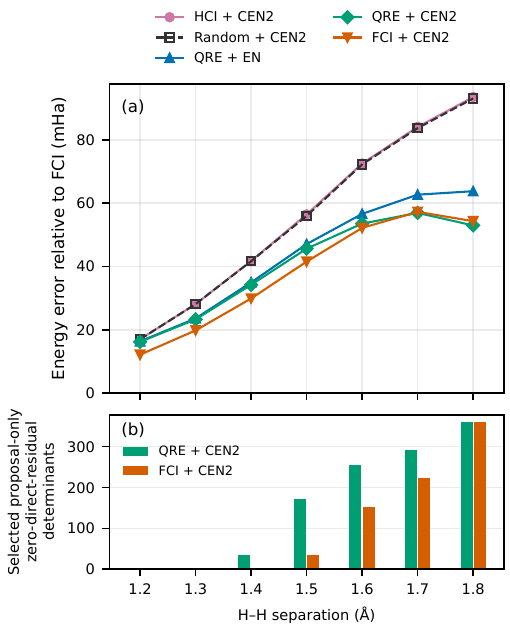}
 \caption{H$_{12}$ one-shot comparison at matched final dimension
 $N=20{,}000$ over 1.2--1.8~\AA. (a) Variational energy error relative to
 FCI for the shared baseline with fixed HCI proposals alone or augmented by
 random, exact-statevector QRE, or exact FCI-distribution proposals, all
 ranked with CEN2;
 QRE+ordinary EN is a ranking ablation.
 The nearly coincident HCI+ordinary-EN curve is omitted.
 The QRE and FCI pools use exact probability ranking after baseline/reference
 exclusion, spatial-occupation aggregation, and spin completion, with the same
 30,000-determinant capacity.
 At 1.2~\AA, the QRE pool contains 12,697 determinants while the FCI pool
 contains 30,000; both reach the cap at the other geometries.
 (b) Selected source-exclusive determinants with zero direct baseline residual
 for QRE+CEN2 and FCI+CEN2.
 The random pool matches the QRE pool size; no random-only zero-residual
 determinants are selected.}
 \label{fig:h12}
\end{figure}

Fig.~\ref{fig:h12}(a) shows that the QRE proposals become more useful as the
chain is stretched. At matched dimension, the improvement of QRE+CEN2 over
HCI+CEN2 grows with bond length. The single random+CEN2 control uses a proposal
pool matched in size to QRE's, yet gives virtually no improvement over
HCI+CEN2. In this comparison, the contrast suggests that the usefulness of
QRE's proposed determinants matters beyond the number of candidates added.

The proposal source matters even with ordinary EN ranking: QRE substantially
lowers the error relative to HCI-only proposals when both pools use the same
ranking criterion. Its quantum proposal state therefore offers an alternative
route to discovering useful determinants absent from the finite HCI proposal
pool. Replacing EN with CEN2 brings a further gain for QRE, especially as the
chain is stretched, but only a small gain of up to 0.44~mHa for HCI-only
proposals. This small gain reflects changes in ranking through couplings
among candidates within the same proposal pool.
QRE+CEN2 achieves comparable performance to FCI+CEN2,
particularly at stretched geometries. FCI proposals give lower errors at
shorter separations, while QRE gives slightly lower errors at the two longest
separations.

The normalized postselected QRE proposal populations have low similarity to
those of the FCI ground state, with squared Bhattacharyya coefficients of
0.30\%--9.44\% (Supplemental Material, Fig.~\ref{fig:supp-fci-overlaps}).
Nevertheless, combining their proposed determinants with classical HCI
candidates improves the final variational energy. QRE therefore supplies
useful complementary determinants without requiring its proposal distribution
to reproduce the FCI populations.

Fig.~\ref{fig:h12}(b) shows that CEN2 selects an increasing number of QRE-only
determinants with zero direct residual against the baseline as the chain is
stretched, whereas ordinary EN and the random+CEN2 control select none.
FCI+CEN2 also selects zero-direct-residual proposal-only determinants,
reaching the same count of 360 as QRE+CEN2 at $R=1.8$~\AA; this diagnostic
is therefore not unique to QRE proposals.
EN gives these determinants zero individual score, whereas CEN2 can admit them through
candidate couplings or a positive-score spin-bundle partner. Their growing
count is consistent with a larger role for indirect pathways at longer bond
lengths, although the count does not isolate their energy contribution or the
benefit of projection. The diagnostic definition and additional energy
comparisons are given in Supplemental Material, Secs.~S1.E and S1.F.

QRE+CEN2 remains lower than HCI+CEN2 after EN-PT2 at all seven geometries.
At $R=1.8$~\AA, an energy advantage of 14.12~mHa remains, compared with
40.51~mHa before correction.

We next examine whether the incremental benefit of QRE persists with finite
measurement samples, including the Reimei demonstration. Repeated outcomes and determinants already present in the baseline or
reference consume shots without enlarging the external pool, reducing the
yield of new candidates as sampling proceeds.

\begin{figure}[h]
 \centering
  \includegraphics[width=\columnwidth]{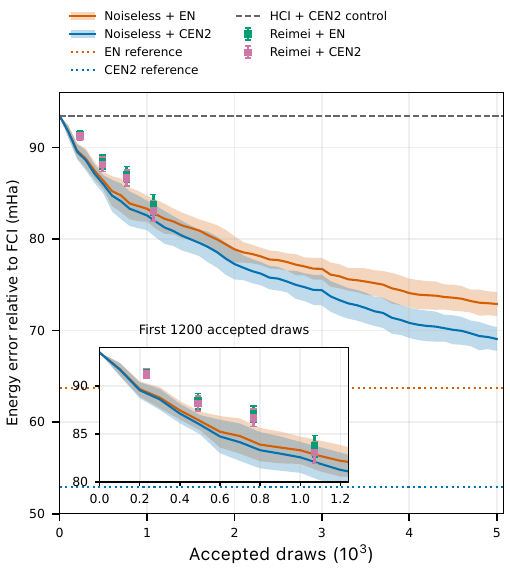}
 \caption{Finite-sampling QRE results for H$_{12}$/STO-3G at $R=1.8$~\AA\
 and final determinant budget $N=20{,}000$. Solid curves show noiseless means
 across ten sampling seeds; shaded bands indicate one sample standard
 deviation. Squares show Reimei results,
 with bars indicating one bootstrap standard deviation conditional on the
 observed samples. Dotted lines show exact-probability QRE references; the
 dashed line is the HCI+CEN2 control. The inset enlarges the first 1,200 accepted
 draws.
 }
 \label{fig:hardware}
\end{figure}

Fig.~\ref{fig:hardware} shows the final variational energy error relative to
FCI as a function of accepted draws at $R=1.8$~\AA. Noiseless samples are drawn
from the full postselected QRE distribution, so repeated outcomes and baseline
or reference hits count toward the draw budget before external-pool
construction. The fixed baseline, classical proposal pool, ranking
procedures, and final determinant budget remain fixed. The dotted lines show
the corresponding exact-probability QRE results from Fig.~\ref{fig:h12}.

On Reimei at $R=1.8$~\AA, four jobs of 2,500 shots each yielded
1,072 accepted draws out of 10,000 attempted shots (10.72\%). These draws
passed both the membership flag and the $N_\alpha=N_\beta=6$ symmetry check
(Table~\ref{tab:hardware-sampling} in End Matter).
The bootstrap procedure and interpretation of the sampling
uncertainty are detailed in End Matter. Ranked with the same HCI pool by CEN2 at the
same final determinant budget, the measured proposals lower the variational
energy relative to classical-only selection, demonstrating that an incremental
benefit remains with finite shots on Reimei. This gain
is smaller than that of the statevector QRE proposals, whereas the larger
random pool gives virtually none. Both implementations use the same
eight-control projector. Hardware reference preparation is approximated to
limit gate cost, and the proposal pool is limited by the accepted measurement
outcomes. These differences, together with device noise, prevent attributing
the energy gap to noise alone.

The hardware results show a modest but positive energy gain from quantum
proposals. The present ansatz explores configurations around a classical
reference and has not been optimized for the yield of useful external
candidates. Because QRE does not require the proposal state to approximate
the target eigenstate, it allows flexibility in circuit structure and
parameters. This freedom provides scope to refine the quantum proposal
ansatz for greater energy improvement within a fixed gate and measurement
budget.

In conclusion, we propose QRE as a quantum determinant-proposal mechanism
that complements a fixed classical SCI baseline through common
energy-based selection and one final variational diagonalization.
In this controlled, noiseless H$_{12}$ comparison at matched final dimension,
adding QRE proposals reduces the FCI error relative to HCI+CEN2 by only
0.81~mHa at $R=1.2$~\AA, but by a substantial 40.51~mHa at $R=1.8$~\AA,
where multireference character is expected to be strongest within the scan.
This increasing contribution is consistent with QRE supplying more useful
complementary determinants as the chain becomes more strongly correlated.
On the Quantinuum Reimei trapped-ion quantum computer, proposals from
finite measurements at $R=1.8$~\AA\ lower the variational energy relative to
HCI+CEN2 at the same final determinant budget.

QRE still requires classical storage and diagonalization of the selected
space, so it does not remove the classical memory and compute limit for SCI.
Quantum phase estimation offers an alternative route to energy estimation
without this final classical diagonalization, but its application to larger
molecular systems is expected to require quantum error correction. An
end-to-end demonstration with error correction has been achieved for molecular
hydrogen in a minimal basis~\cite{Yamamoto2026QEC}; extending this approach
to more complex molecules remains a substantial challenge.
Future studies should test whether
the trends observed here extend to other molecular systems and correlation
regimes, and how the useful proposal yield depends on the classical baseline
and quantum ansatz. Such comparisons will clarify where QRE can make limited
classical CI memory more effective in practice.

\begin{acknowledgments}
We thank Luca Erhart, Silas Dilkes, Carlo A. Gaggioli, and
Yuri Alexeev for useful discussions.
We thank Nathan Lysne, Craig Holliman, Tomohiro Murata, and Yuga
Kodama for their support in executing the circuits on Reimei.
Part of this study is based on results obtained from a project, JPNP20017,
commissioned by the New Energy and Industrial Technology Development
Organization (NEDO).
W.D. acknowledges the FOCUS (Hyogo Prefecture, Kobe City) Project to Promote
the Formation of a Research and Education Center (COE) in the Field of
Computational Science.
OpenAI Codex assisted with research and manuscript preparation, as described
in End Matter.
\end{acknowledgments}

\bibliography{main}

\clearpage
\twocolumngrid
\section*{End Matter}

\textit{Evolutionary algorithm reference preparation---} We approximate a target reference state $\ket{\Phi^R}$, specified by CI coefficients, using a preparation circuit $U_C(\boldsymbol{\theta})$ acting on an initial state $\ket{\Phi_\mathrm{in}}$.
For the hardware demonstration, we use the evolutionary-search component of
the Evolutionary Agent framework~\cite{Yuri} to construct the
reference-preparation circuit classically.
The circuit structure is an ordered sequence of parameterized single- and
double-excitation blocks. The search targets a squared overlap of at least
0.95 with the reduced reference while favoring lower gate counts among
circuits satisfying this threshold. Candidates satisfying this threshold are preferred over
those that do not. Below the threshold, candidates are ranked by fidelity;
among candidates satisfying it, lower two-qubit gate counts are preferred.

An evolutionary search explores circuit structures and parameter initializations.
Tournament selection chooses parent circuits.
New candidates are generated from these parents by block insertion, deletion, random angle perturbation.%

Each candidate inherits unmodified angles from its parent.
All free angles are then refined using a derivative-free optimizer.
Fidelity is evaluated by noiseless statevector simulation.
At each generation, circuits are selected according to this
threshold-prioritized ranking.
The best circuit found during the search is returned.
The evolutionary algorithm optimization for the hardware reference-preparation
circuit was executed on the ROQUO GPU supercomputer.

\label{app:hardware-specifications}%
\textit{Quantinuum Reimei specifications---} Quantinuum Reimei is a
56-qubit System Model H2 trapped-ion quantum
computer~\cite{QuantinuumReimei2026}.
Table~\ref{tab:reimei-specifications} summarizes the device
specifications~\cite{QuantinuumPerformanceValidation2026}.

\begin{table}[b]
\caption{Reimei specifications dated 2026-09-10.}
\label{tab:reimei-specifications}
\begin{ruledtabular}
\begin{tabular}{@{}p{0.48\columnwidth}p{0.46\columnwidth}@{}}
Characteristic & Value \\
\hline
Physical Qubits & 56 \\
1-Qubit Gate Error ($P_1$) & $(2.80 \pm 0.50)\times10^{-5}$ \\
2-Qubit Gate Error ($P_2$) & $(1.20 \pm 0.07)\times10^{-3}$ \\
SPAM Error ($P_{\mathrm{Meas}\,0}$) & $(7.00 \pm 0.90)\times10^{-4}$ \\
SPAM Error ($P_{\mathrm{Meas}\,1}$) & $(4.00 \pm 0.20)\times10^{-3}$ \\
Mid-Circuit Measurement Crosstalk Error & $(2.00 \pm 0.04)\times10^{-5}$ \\
Memory (Idle) Error & $(2.80 \pm 0.30)\times10^{-4}$ \\
\end{tabular}
\end{ruledtabular}
\end{table}

\textit{Circuit implementation and compilation---}
The noiseless benchmark circuit is initialized directly
from the normalized truncated-reference coefficients.
It applies ten generators in descending order of the score in
Eq.~\eqref{eq:generator-score}, with a common angle of $0.15$~rad.
An eight-control $Q$ projection follows the fifth generator, giving a
5:5 split. It checks all eight frozen spin orbitals---four occupied core
and four empty virtual orbitals---and rejects configurations matching their
reference occupations. The accepted state is normalized after
projection. The generator list, projector definition, and acceptance
probabilities are given in Supplemental Material, Sec.~S1.D.

The hardware circuit acts on 31 qubits: 24 system qubits, one per spin orbital under the Jordan--Wigner mapping, six work qubits, and one flag qubit.
On the system register, $X$ gates initialize the four frozen core spin orbitals as occupied, and a 16-qubit circuit prepares the reference in the CAS(8,8) window.
The ten transport generators are then applied as products of Pauli-string exponentials, one spin component at a time, so that particle number and $S_z$ are conserved exactly.
The projector identifies the reference occupation pattern of the eight frozen spin orbitals by evaluating a logical AND, so that $Q$ excludes the full embedded-reference sector.
After $X$ gates map the reference occupation pattern to all ones, the eight-controlled NOT is decomposed into Toffoli gates with work qubits~\cite{Barenco1995}.
These gates form a balanced tree, storing intermediate ANDs in six clean work qubits and the final result in the flag qubit.
Intermediate nodes use relative-phase Toffoli gates, with three rather than six CNOTs; their phases cancel because each is undone by its inverse before the flag is measured~\cite{Maslov2016}.

Each component is optimized separately with pytket~\cite{Sivarajah2020}, and no compilation pass acts across the projector.
The reference circuit, built from Givens-type and paired-double gates, is optimized by FullPeepholeOptimise without wire swaps.
The early and late generator blocks are each resynthesized by GreedyPauliSimp~\cite{Schmitz2024GraphOptimization} (20 trials).
The assembled circuit is then compiled by the Reimei pass at optimization level 2 into native PhasedX, $R_z$, and ZZPhase gates.
Dynamical decoupling is added by the device-side compiler at submission.
The compiled circuit contains 939 ZZPhase gates.

\textit{Measurement processing and candidate construction---} Each measured
occupation string is checked against the target particle numbers $N_\alpha$
and $N_\beta$, after membership-flag postselection when applicable. Accepted
strings are deduplicated, and determinants already present in the protected
baseline $\mathcal V_0$ are excluded from the external proposal pool.
Repeated outcomes and baseline hits still count toward the measurement-shot
budget.

Each remaining string is mapped to its spatial occupation. Spin completion
generates all assignments of the unpaired spins consistent with the target
$M_S=(N_\alpha-N_\beta)/2$, forming a complete spin bundle. These generated
partners are candidates, not additional measured outcomes. Any proposal-pool
cap retains complete bundles under the benchmark-specific rules in
Supplemental Material, Sec.~S1.D.

The quantum set $\mathcal Q$ contains the retained bundles and reference
support $\mathcal R\setminus\mathcal V_0$, when used. Combining it with the
classical pool gives
$\mathcal C=\mathcal H\cup\mathcal Q$. Measurement frequencies do not enter
the subsequent EN/CEN2 scores or provide a reserved quota for quantum
candidates. The protected baseline remains in the final space, and complete
bundles compete under the common energetic ranking.

\textit{Finite-sampling analysis---} The four Reimei jobs summarized in
Table~\ref{tab:hardware-sampling} each contain 2,500 attempted shots.
The respective numbers of draws passing both the membership flag and
$N_\alpha=N_\beta=6$ symmetry check are 234, 256, 277, and 305.
Combining jobs in order gives cumulative accepted-sample counts of
234, 490, 767, and 1,072. At each checkpoint, accepted determinants are
deduplicated and spin-completed before baseline/reference exclusions.

\begin{table}[b]
\caption{Reimei sampling statistics at $R=1.8$~\AA.
Flag is the fraction of attempted shots with $m_f=0$; symmetry is the
fraction of flag-accepted shots satisfying $N_\alpha=N_\beta=6$;
usable is the fraction of attempted shots passing both checks.
Draws counts accepted outcomes passing both checks, including repeats. }
\label{tab:hardware-sampling}
\small
\setlength{\tabcolsep}{3pt}
\begin{ruledtabular}
\begin{tabular}{c r r r r r}
Job & Shots & Flag (\%) & Symmetry (\%) & Usable (\%) & Draws\\
\hline
1 & 2,500 & 71.8 & 13.0 & 9.4 & 234\\
2 & 2,500 & 58.4 & 17.5 & 10.2 & 256\\
3 & 2,500 & 67.5 & 16.4 & 11.1 & 277\\
4 & 2,500 & 67.0 & 18.2 & 12.2 & 305\\
\end{tabular}
\end{ruledtabular}
\end{table}

For each of 100 bootstrap replicates, we resample each job independently
with replacement from its accepted outcomes, preserving that job's accepted-shot
count. Resampled jobs are combined at each cumulative checkpoint, and the
same resample is used for EN and CEN2. We repeat deduplication, spin
completion, exclusions, energetic ranking, selection, and final
diagonalization for each replicate. The baseline, HCI proposal pool, and
final dimension $N=20{,}000$ remain fixed. Hardware markers show the energies
from the original samples; error bars show one sample standard deviation across
bootstrap replicates.
In these data, bootstrap resampling reduces candidate diversity and lowers
the mean gain relative to the original sample. The error bars describe
sensitivity to resampling the observed outcomes, rather than confidence
intervals or between-run variability.

\textit{Perturbative energy correction---} We evaluate EN-PT2~\cite{Huron1973}
for the final $N=20{,}000$ HCI+CEN2 and QRE+CEN2 states at all seven geometries
in the statevector benchmark. For the normalized final CI coefficients
$C_J^{\mathrm{var}}$ and variational energy $E_{\mathrm{var}}$, the correction is
\begin{equation}
\begin{aligned}
 E^{(2)} &= \sum_{I\notin\mathcal V_{\mathrm{final}}}
 \frac{\left|\sum_{J\in\mathcal V_{\mathrm{final}}}
 H_{IJ}C_J^{\mathrm{var}}\right|^2}{E_{\mathrm{var}}-H_{II}},\\
 E_{\mathrm{corr}} &= E_{\mathrm{var}}+E^{(2)}.
\end{aligned}
\label{eq:en-pt2}
\end{equation}
The residuals use the final eigenvector, rather than the initial HCI
baseline or the CEN2 ranking amplitudes. A matrix-free Hamiltonian action
evaluates them over the full $N_\alpha=N_\beta=6$ sector of determinants.
The main-text figures show the uncorrected variational values; the EN-PT2
comparison in Supplemental Material, Fig.~\ref{fig:supp-pt2}, is a separate
diagnostic of the selected spaces.

\textit{Software---} The validated prototype analysis environment uses
Python 3.13.12, PySCF 2.14.0, NumPy 2.5.1, SciPy 1.18.0, and Matplotlib 3.11.2.
PySCF~\cite{Sun2018} supplies molecular integrals and
reference electronic-structure calculations. Custom selection and statevector
routines use NumPy~\cite{Harris2020} for array operations and
SciPy~\cite{Virtanen2020} for sparse linear algebra and eigensolves;
Matplotlib generates the figures. Circuit angles in the evolutionary search
are refined using COBYLA~\cite{Powell1994}, while CUDA-Q~\cite{cudaq} evaluates
reference-state fidelities by statevector simulation.%
Circuit compilation and optimization use
pytket~\cite{Sivarajah2020}.
The hardware workflow uses Tierkreis~\cite{Sivarajah2022} for
orchestration and Quantinuum Nexus~\cite{QuantinuumNexus2024} for job submission
and result retrieval.

\textit{AI-assisted research tools---} OpenAI Codex (model
\texttt{gpt-6.1-sol}, accessed 23 October 2026) assisted with literature
synthesis, scientific discussion and interpretation, manuscript drafting
and revision, and developing and debugging analysis and plotting scripts
under the authors' direction.
The authors retain responsibility for the scientific content and conclusions.

\clearpage
\onecolumngrid
\setcounter{equation}{0}
\renewcommand{\theequation}{S\arabic{equation}}
\renewcommand{\theHequation}{supp.\arabic{equation}}
\setcounter{table}{0}
\renewcommand{\thetable}{S\arabic{table}}
\renewcommand{\theHtable}{supp.\arabic{table}}
\setcounter{figure}{0}
\renewcommand{\thefigure}{S\arabic{figure}}
\renewcommand{\theHfigure}{supp.\arabic{figure}}

\begin{center}
 {\large\bfseries Supplemental Material for\\[2pt]
 ``Quantum Residual Exploration for Selected Configuration Interaction''}
 \vspace{6pt}

 Kentaro Yamamoto, Ikko Hamamura, Juan W. Pedersen, Duncan Gowland,
 Gabriel Laude, Shintaro Maekawa, David Mu\~{n}oz Ramo, Elica Kyoseva, and William Dawson
\end{center}

\section*{S1. Workflow implementation and statevector benchmark}

We describe the construction of the common HCI baseline, the classical and
quantum proposal pools, and their selection into the final CI space.
Comparisons use the same final determinant dimension; total computational
costs are not matched. Section~S2 examines additional HCI discovery;
hardware methods and analysis are given in End Matter.

\subsection*{S1.A. Electronic Hamiltonian construction}

The calculations use an
equally spaced linear H$_{12}$ chain in the STO-3G basis.  All 12 electrons
and 12 spatial orbitals are correlated in a fixed-$M_S=0$ determinant basis.
FCI in this CAS(12,12) space supplies exact reference energies and the
explicit FCI+CEN2 proposal-source control. It does not enter QRE proposal
generation or the determinant-ranking scores.

\begin{figure}[htbp]
 \centering
 \includegraphics[width=0.65\textwidth]{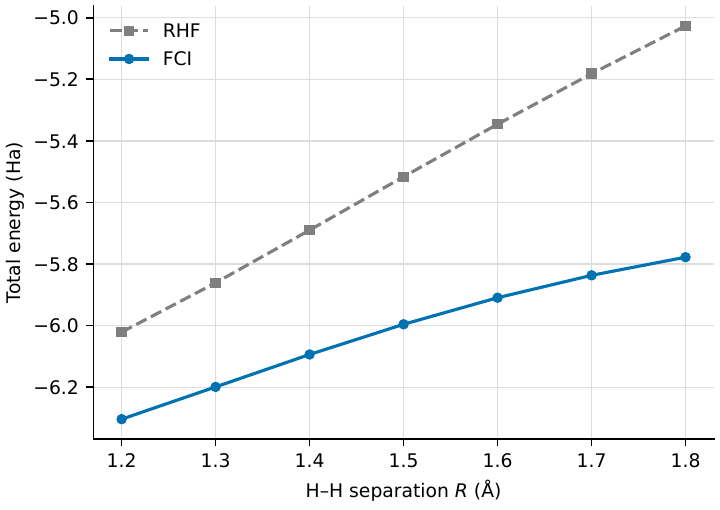}
 \caption{RHF and FCI total energies for the equally spaced linear
 H$_{12}$ chain in STO-3G.}
 \label{fig:supp-hf-fci}
\end{figure}

The Hamiltonian is expressed in canonical restricted-Hartree--Fock (RHF)
orbitals. Fig.~\ref{fig:supp-hf-fci} compares the RHF and FCI total energies
at the seven benchmark separations from 1.2 to 1.8~\AA; both include nuclear
repulsion. The energy lowering from RHF to FCI grows from 0.28 to 0.75~Ha
over this range, illustrating the increasing correlation energy missing
from the single-determinant RHF description as the chain is stretched.

\subsection*{S1.B. Protected HCI baseline}

At each geometry, we order the determinants of the centered CAS(8,8)
ground state by their squared coefficients and retain enough to capture at
least 99\% of its probability weight. We then complete the spin configurations
and normalize the truncated state. This 185--798-determinant reference
initializes QRE and seeds the HCI baseline.

HCI expands this initial space by adding all determinants that satisfy the
threshold $\epsilon_0=2.3$~mHa, together with their spin partners. After each
expansion, the Hamiltonian is rediagonalized to obtain the lowest singlet.
Expansion stops when a selection step adds fewer than 1\% of the preceding
number of determinants. These last additions are retained, and a final
diagonalization defines the fixed baseline. The threshold was chosen in
pilot calculations and held fixed across geometries; the baseline grows
without a determinant cap.

The HCI stopping criterion defines the classical starting point for this
comparison. The benefit of QRE depends on the complementary candidates it
supplies relative to the classical proposal pool, rather than on baseline
dimension alone.

\begin{figure}[htbp]
 \centering
 \includegraphics[width=0.90\textwidth]{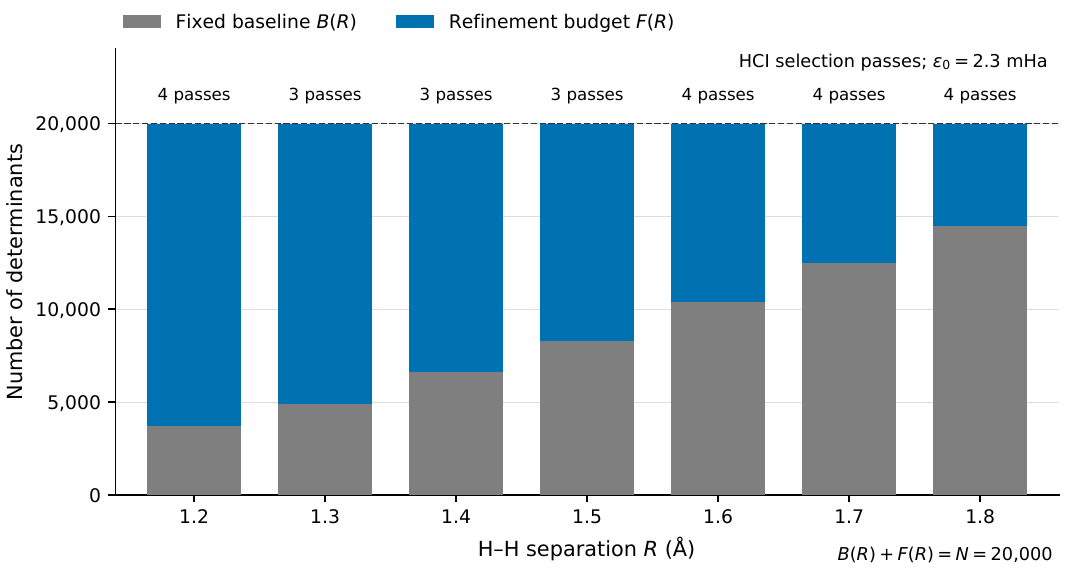}
 \caption{Fixed HCI baseline and one-shot refinement budgets for
 H$_{12}$/STO-3G. Stacked bars show the baseline dimension $B(R)$ and
 refinement budget $F(R)$, totaling $N=20{,}000$ at every geometry.
 Annotations above the bars give the number of HCI selection passes,
 including the stopping pass.
 A final diagonalization follows. All geometries use $\epsilon_0=2.3$~mHa.}
 \label{fig:supp-budgets}
\end{figure}

Fig.~\ref{fig:supp-budgets} shows the resulting baseline dimensions $B(R)$
and the available refinement budgets $F(R)=20{,}000-B(R)$. In the one-shot
benchmark, all branches retain this baseline and select additions to reach
the common final dimension $N=20{,}000$. The baseline energy errors in
Fig.~\ref{fig:supp-hci-energies} provide the starting point for these refinements.

\subsection*{S1.C. Classical proposals}

Starting from the fixed baseline, we construct a 30,000-determinant HCI
proposal pool using the score in Eq.~\eqref{eq:hci}, evaluated with the
baseline coefficients $C_J^{(0)}$. Scores are summed within complete spin
bundles. To allow the final space to contain exactly $N=20{,}000$
determinants without splitting a bundle, we reserve the two highest-ranked
closed-shell determinants, each of which forms a one-determinant spin
bundle. The remaining 29,998 places follow the HCI bundle ranking.
This fixed HCI pool is shared by the compared proposal branches.

\begin{figure}[htbp]
 \centering
 \includegraphics[width=0.65\textwidth]{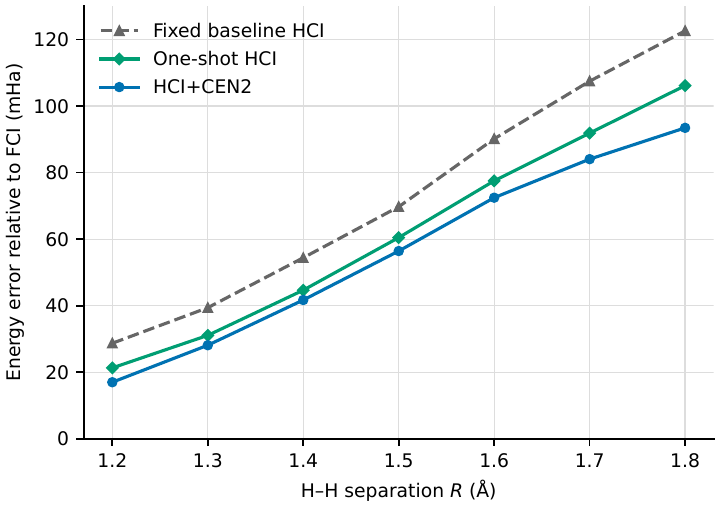}
 \caption{Variational energy errors relative to FCI for fixed-baseline HCI,
 one-shot HCI, and HCI+CEN2 for the equally spaced linear H$_{12}$ chain in
 STO-3G. The fixed baseline uses
 $\epsilon_0=2.3$~mHa. Both refinements use the same classical proposal pool
 and final dimension $N=20{,}000$; one-shot HCI uses the original HCI
 ranking, while HCI+CEN2 reranks candidates using CEN2.}
 \label{fig:supp-hci-energies}
\end{figure}

To assess refinement using the original HCI ranking, we select $F(R)$
additions from this pool using the same HCI bundle scores and diagonalize
the resulting space once. Fig.~\ref{fig:supp-hci-energies} compares this
one-shot HCI result with HCI+CEN2, which reranks the same candidates.
At $R=1.8$~\AA, the original HCI refinement lowers the baseline energy by
16.46~mHa, compared with 29.16~mHa using CEN2 at the same final dimension.

For the random control, the HCI pool is augmented by a random pool matching
the QRE pool size. Determinants are sampled uniformly without replacement
from the $N_\alpha=N_\beta=6$ sector and spin-completed. A single random pool
is used at each geometry, so the control does not characterize variability
across random pools.
Random candidates are not preselected by their CEN2 scores. They are
combined with the common HCI pool and ranked by the same CEN2 procedure
as the QRE candidates. This control tests whether enlarging the candidate
pool through unguided sampling provides a comparable benefit.

\subsection*{S1.D. Quantum proposals}

The circuit starts from the normalized reference described in Sec.~S1.B,
embedded in the target CAS(12,12) space. Any reference determinants outside
the fixed baseline are included in all compared candidate pools. Reference
amplitudes guide the choice of transport generators, while the baseline
coefficients $C_J^{(0)}$ and energy $E_0$ determine the final EN/CEN2 scores.

The transport generators are $G_\mu=T_\mu-T_\mu^\dagger$, with $\mu$ labeling
the orbital indices and excitation type. Elementary single and double terms
in $T_\mu$ are $a_p^\dagger a_q$ and
$a_p^\dagger a_q^\dagger a_s a_r$, respectively; $a_p^\dagger$ creates and
$a_p$ annihilates an electron in spin orbital $p$. Spin components associated
with the same spatial-orbital excitation are combined. The anti-Hermitian
generator produces unitary transport $e^{\theta_\mu G_\mu}$ for real
$\theta_\mu$. The generator-ranking score $w_\mu$ is defined in
Eq.~\eqref{eq:generator-score}; its EN denominator uses the numerical floor
specified in Sec.~S1.E.

Generator selection is a classical preprocessing step with a computational
cost comparable to an EN-PT2 evaluation over the relevant external space,
together with the additional cost of evaluating reference-accessibility
amplitudes and accumulating generator scores.

We apply the ten highest-scoring spin-adapted, particle-conserving
generators in descending order of $w_\mu$, with the common angle
$\theta_\mu=0.15$~rad at every geometry. Spin components enter $T_\mu$ with
unit coefficients; neither generators nor their external actions are
rescaled to unit norm.

Table~\ref{tab:supp-transport-operators} lists the ten selected transport
generators and their ranking scores at every benchmark geometry.
Both singles and doubles are eligible under the same ranking criterion,
but singles rank below the ten selected doubles at every geometry.
The notation $(i,j)\to(a,b)$ denotes
\begin{equation}
 T_{ab\leftarrow ij}=\sum_{\sigma,\tau\in\{\alpha,\beta\}}
 a^\dagger_{a\sigma}a^\dagger_{b\tau}a_{j\tau}a_{i\sigma},
 \qquad G_{ab\leftarrow ij}=T_{ab\leftarrow ij}-T_{ab\leftarrow ij}^\dagger.
\end{equation}
Here $i,j,a,b\in\{0,\ldots,11\}$ are zero-based canonical spatial-orbital
indices; spin orbitals $p\alpha$ and $p\beta$ correspond to indices $2p$
and $2p+1$, respectively. Repeated spatial indices denote pair excitations.
The tabulated orientations and order are retained as used in the circuit,
including inverse pairs and tied scores.

The selected scores tend to become larger as the chain is stretched,
spanning $(1.80$--$2.93)\times10^{-5}$~Ha at $R=1.2$~\AA\ and
$(8.13$--$21.53)\times10^{-5}$~Ha at $R=1.8$~\AA.
These larger scores indicate greater combined reference accessibility and
baseline EN importance of the external determinants reached by the selected
generators. The operator identities and their ordering also vary with geometry.

\begin{table}[htbp]
 \centering
 \caption{Ordered transport generators and ranking scores for the statevector
 benchmark. Each cell gives the spatial-orbital excitation
 $(i,j)\to(a,b)$ and, beneath it, $w_\mu$ from Eq.~\eqref{eq:generator-score}
 in $10^{-6}$~Ha. Column headings give $R$ in \AA. Rows give the executed
 order; the horizontal rule after row five marks the eight-control $Q$ projection.
 Orbital indices start at zero.}
 \label{tab:supp-transport-operators}
 \scriptsize
 \setlength{\tabcolsep}{4pt}
 \renewcommand{\arraystretch}{1.3}
 \begin{tabular}{c c c c c c c c}
 \hline\hline
 $\mu$ & 1.2 & 1.3 & 1.4 & 1.5 & 1.6 & 1.7 & 1.8\\
 \hline
 1 & \shortstack{$(0,2)\!\to\!(9,11)$\\$29.343$} & \shortstack{$(0,1)\!\to\!(10,11)$\\$37.061$} & \shortstack{$(1,2)\!\to\!(9,10)$\\$59.102$} & \shortstack{$(0,1)\!\to\!(10,11)$\\$61.527$} & \shortstack{$(0,1)\!\to\!(10,11)$\\$129.905$} & \shortstack{$(0,1)\!\to\!(10,11)$\\$183.969$} & \shortstack{$(0,1)\!\to\!(10,11)$\\$215.290$}\\
 2 & \shortstack{$(0,3)\!\to\!(8,11)$\\$23.890$} & \shortstack{$(1,2)\!\to\!(9,10)$\\$32.509$} & \shortstack{$(0,1)\!\to\!(10,11)$\\$51.446$} & \shortstack{$(0,2)\!\to\!(9,11)$\\$57.618$} & \shortstack{$(1,2)\!\to\!(9,10)$\\$103.729$} & \shortstack{$(0,2)\!\to\!(9,11)$\\$140.517$} & \shortstack{$(1,2)\!\to\!(9,10)$\\$181.883$}\\
 3 & \shortstack{$(0,1)\!\to\!(10,11)$\\$23.427$} & \shortstack{$(0,3)\!\to\!(8,11)$\\$30.832$} & \shortstack{$(0,2)\!\to\!(9,11)$\\$47.232$} & \shortstack{$(1,3)\!\to\!(8,10)$\\$53.783$} & \shortstack{$(0,3)\!\to\!(8,11)$\\$95.145$} & \shortstack{$(1,1)\!\to\!(10,10)$\\$102.198$} & \shortstack{$(0,2)\!\to\!(9,11)$\\$144.590$}\\
 4 & \shortstack{$(8,9)\!\to\!(2,3)$\\$22.817$} & \shortstack{$(1,3)\!\to\!(8,10)$\\$27.024$} & \shortstack{$(1,3)\!\to\!(8,10)$\\$45.437$} & \shortstack{$(0,0)\!\to\!(11,11)$\\$52.225$} & \shortstack{$(1,3)\!\to\!(8,10)$\\$88.917$} & \shortstack{$(1,3)\!\to\!(8,10)$\\$95.550$} & \shortstack{$(1,3)\!\to\!(8,10)$\\$141.075$}\\
 5 & \shortstack{$(2,3)\!\to\!(8,9)$\\$22.817$} & \shortstack{$(0,4)\!\to\!(7,11)$\\$25.429$} & \shortstack{$(1,4)\!\to\!(7,10)$\\$39.533$} & \shortstack{$(1,2)\!\to\!(9,10)$\\$48.610$} & \shortstack{$(0,2)\!\to\!(9,11)$\\$85.271$} & \shortstack{$(0,3)\!\to\!(8,11)$\\$94.245$} & \shortstack{$(0,3)\!\to\!(8,11)$\\$126.651$}\\
 \hline
 6 & \shortstack{$(9,9)\!\to\!(2,2)$\\$20.248$} & \shortstack{$(0,2)\!\to\!(9,11)$\\$24.806$} & \shortstack{$(0,3)\!\to\!(8,11)$\\$36.565$} & \shortstack{$(1,1)\!\to\!(10,10)$\\$47.607$} & \shortstack{$(1,1)\!\to\!(10,10)$\\$68.705$} & \shortstack{$(0,4)\!\to\!(7,11)$\\$90.474$} & \shortstack{$(1,4)\!\to\!(7,10)$\\$111.473$}\\
 7 & \shortstack{$(2,2)\!\to\!(9,9)$\\$20.248$} & \shortstack{$(1,4)\!\to\!(7,10)$\\$23.278$} & \shortstack{$(1,1)\!\to\!(10,10)$\\$31.057$} & \shortstack{$(0,4)\!\to\!(7,11)$\\$45.006$} & \shortstack{$(0,4)\!\to\!(7,11)$\\$64.812$} & \shortstack{$(1,2)\!\to\!(9,10)$\\$89.166$} & \shortstack{$(1,1)\!\to\!(10,10)$\\$97.941$}\\
 8 & \shortstack{$(8,8)\!\to\!(3,3)$\\$18.863$} & \shortstack{$(1,1)\!\to\!(10,10)$\\$19.594$} & \shortstack{$(0,4)\!\to\!(7,11)$\\$29.614$} & \shortstack{$(0,3)\!\to\!(8,11)$\\$44.067$} & \shortstack{$(0,5)\!\to\!(6,11)$\\$57.876$} & \shortstack{$(0,5)\!\to\!(6,11)$\\$82.237$} & \shortstack{$(1,5)\!\to\!(6,10)$\\$93.226$}\\
 9 & \shortstack{$(3,3)\!\to\!(8,8)$\\$18.863$} & \shortstack{$(0,0)\!\to\!(11,11)$\\$15.426$} & \shortstack{$(0,0)\!\to\!(11,11)$\\$24.410$} & \shortstack{$(1,4)\!\to\!(7,10)$\\$36.265$} & \shortstack{$(1,4)\!\to\!(7,10)$\\$57.044$} & \shortstack{$(0,0)\!\to\!(11,11)$\\$76.981$} & \shortstack{$(0,4)\!\to\!(7,11)$\\$87.665$}\\
 10 & \shortstack{$(1,2)\!\to\!(9,10)$\\$17.963$} & \shortstack{$(1,4)\!\to\!(6,9)$\\$15.030$} & \shortstack{$(9,9)\!\to\!(2,2)$\\$21.312$} & \shortstack{$(1,5)\!\to\!(6,10)$\\$32.109$} & \shortstack{$(1,5)\!\to\!(6,10)$\\$54.643$} & \shortstack{$(1,4)\!\to\!(7,10)$\\$74.844$} & \shortstack{$(0,0)\!\to\!(11,11)$\\$81.297$}\\
 \hline\hline
 \end{tabular}
\end{table}

After the fifth generator, an eight-control $Q$ projection rejects the
reference occupation pattern on all frozen spin orbitals, giving a 5:5
split between early and late evolution.
Let $\mathcal F_P=\{0,1,2,3,20,21,22,23\}$ denote the frozen spin-orbital
controls. Their reference occupations $\bar n_j$ are one for indices 0--3
and zero for indices 20--23.
The projectors and flag oracle are
\begin{equation}
 P=\prod_{j\in\mathcal F_P}|\bar n_j\rangle\!\langle\bar n_j|,
 \qquad Q=I-P,\qquad
 \mathcal O_P=P\otimes X_f+Q\otimes I_f,
 \label{eq:membership-oracle}
\end{equation}
with identities on unselected qubits implicit. Here $X_f$ flips the flag
qubit and $I_f$ is its identity. For an early-block state $\ket{\psi}$ and
flag initialized to $\ket{0}_f$,
\begin{equation}
 \mathcal O_P\ket{\psi}\ket{0}_f
 =P\ket{\psi}\ket{1}_f+Q\ket{\psi}\ket{0}_f.
\end{equation}
Retaining flag outcome $m_f=0$ gives $Q\ket{\psi}/\sqrt{p_Q}$ with probability
$p_Q=\|Q\ket{\psi}\|^2$. The ordered transport gates and split after gate
$k$ are shown in Eq.~\eqref{eq:projected-cre}. This projector excludes the full
embedded-reference sector. Rejected outcomes avoid late evolution only when conditional early
termination is implemented; the executed hardware circuit does not use it.

The projection acceptance probabilities $p_Q$ are listed in
Table~\ref{tab:supp-acceptance}. They range from 0.2338 to 0.3491 over the seven
geometries.  Thus a retained projected statevector sample corresponds to an
ideal acceptance overhead $1/p_Q$ before accounting for preparation cost,
duplicate samples, symmetry rejection, or device noise.

\begin{table}[htbp]
 \centering
 \caption{Ideal projection acceptance for H$_{12}$/STO-3G.
 The probability $p_Q$ is the squared norm retained by the eight-control
 $Q$ projection after the fifth of ten transport generators.
 Values describe the noiseless benchmark.}
 \label{tab:supp-acceptance}
 \begin{tabular}{c r r r r r r r}
 \hline\hline
 $R$ (\AA) & 1.2 & 1.3 & 1.4 & 1.5 & 1.6 & 1.7 & 1.8\\
 \hline
 $p_Q$ & 0.23 & 0.34 & 0.34 & 0.35 & 0.35 & 0.35 & 0.34\\
 \hline\hline
 \end{tabular}
\end{table}

To construct the QRE proposal pool, we group the exact postselected
probabilities by spatial occupation and exclude the baseline and reference
support. Each remaining occupation is assigned its total probability and
completed with all spin configurations compatible with $M_S=0$. These
complete bundles are admitted in descending probability order up to the
30,000-determinant capacity. The resulting pool contains 12,697 determinants
at 1.2~\AA\ and reaches the capacity at all other geometries. These
probabilities determine pool construction only; final admission is decided
by EN or CEN2 ranking, with no reserved quota for quantum proposals.

For the FCI proposal control, exact FCI probabilities replace the QRE
probabilities in this pool-construction procedure, with the same exclusions,
spatial-occupation aggregation, spin completion, and capacity. FCI amplitudes
do not enter the energetic ranking scores.

\begin{figure}[htbp]
 \centering
 \includegraphics[width=0.80\textwidth]{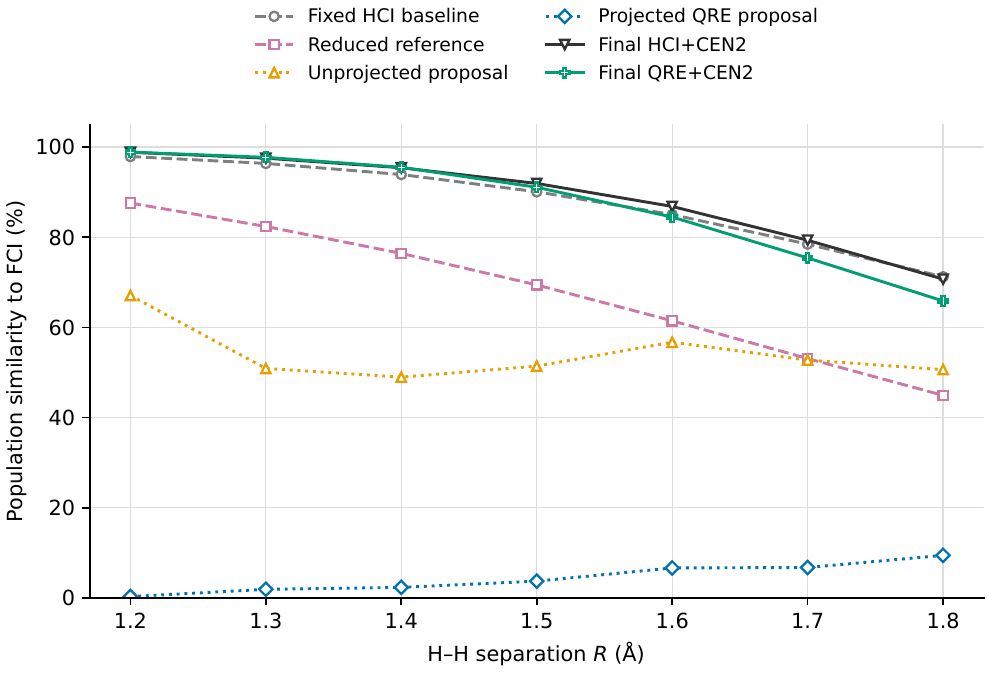}
 \caption{Determinant-population similarity to FCI, measured by
 $100F_{\mathrm{pop}}$, for the noiseless
 benchmarks. The reduced reference is the normalized truncated CAS(8,8)
 input. The unprojected proposal uses the same reference, transport
 generators, angles, and ordering as QRE, with the intermediate projection
 omitted. Final states are obtained by CI diagonalization in the selected
 spaces; final QRE+CEN2 results use projected proposals. Lines connect the
 calculated geometries as guides to the eye.}
 \label{fig:supp-fci-overlaps}
\end{figure}

Fig.~\ref{fig:supp-fci-overlaps} compares determinant populations using the
squared Bhattacharyya coefficient
\begin{equation}
 F_{\mathrm{pop}}(X)=\left(\sum_I\sqrt{p_I^{\mathrm{FCI}}p_I^X}\right)^2,
 \qquad p_I^X=|\langle D_I|\Psi_X\rangle|^2.
\end{equation}
Each distribution is normalized over the full determinant space; the QRE
proposal distribution is conditional on successful projection. This measure
ranges from zero to one and excludes relative-phase information. It is a
post hoc diagnostic and does not enter the QRE calculation.

The projected QRE proposal populations have low similarity to FCI, with
$F_{\mathrm{pop}}$ increasing from 0.30\% to 9.44\% across the scan.
Nevertheless, their determinants complement the classical HCI candidates:
at $R=1.8$~\AA, QRE+CEN2 reduces the energy error from 93.45 to 52.95~mHa
relative to HCI+CEN2. Thus, a proposal distribution need not reproduce FCI
populations to improve the selected-space energy after classical
rediagonalization. Omitting projection increases population similarity at
every geometry, but this alone does not establish which proposal source
would yield a better final selected space.

We use a fixed common transport angle rather than optimizing individual
angles. To examine local sensitivity, we compare 0.10, 0.15, and 0.20~rad
at $R=1.8$~\AA, retaining the same reference, generator sequence,
projector, classical baseline, candidate-pool construction rules, and
final dimension $N=20{,}000$. Candidate pools are constructed from exact
statevector probabilities.

Table~\ref{tab:supp-angle-sensitivity} shows that the QRE+CEN2 energy
improvement changes by only 0.57~mHa across these settings. The 0.10- and
0.20-rad calculations select identical final determinant spaces, giving
the same energy. Acceptance and external-candidate probability vary more
strongly, so this energy stability does not imply equal finite-shot
efficiency.

\begin{table}[htbp]
 \centering
 \caption{Sensitivity to the common transport angle at $R=1.8$~\AA.
 External probability is the probability of measuring a determinant outside
 the fixed HCI baseline, conditional on projection acceptance. Energy
 improvements are relative to HCI+CEN2 at the same final dimension
 $N=20{,}000$.}
 \label{tab:supp-angle-sensitivity}
 \begin{tabular}{c c c c}
 \hline\hline
 Common angle (rad) & Acceptance $p_Q$ &
 \shortstack{External probability\\given acceptance} &
 \shortstack{Energy improvement\\(mHa)}\\
 \hline
 0.10 & 0.1670 & 0.1582 & 41.08\\
 0.15 & 0.3391 & 0.2138 & 40.51\\
 0.20 & 0.5240 & 0.2846 & 41.08\\
 \hline\hline
 \end{tabular}
\end{table}

To examine sensitivity to the input state, we repeat the statevector
calculation at $R=1.8$~\AA\ starting from the RHF determinant instead of
the truncated CAS(8,8) reference. We retain the original CAS-reference-ranked
generators, their order and angles, the eight-control projection after the
fifth generator, the fixed HCI baseline and classical proposal pool, and
the final dimension $N=20{,}000$. Candidate pools are constructed from
exact probabilities using the same spin-completion and selection rules.

With CEN2 ranking, the HF input gives an FCI energy error of 78.34~mHa,
improving on the classical-only error of 93.45~mHa by 15.11~mHa. The
correlated input gives a larger improvement of 40.51~mHa. The HF input
also reduces the accepted probability outside the baseline from 21.38\%
to 2.68\%, indicating a potential sampling-efficiency tradeoff. Thus, a
simple input state can supply useful proposals through the chosen transport
circuit, while the correlated reference improves their effectiveness in
this example. This test retains classical CAS information in the generator
selection and baseline construction.

\textit{Early termination---} Conditional termination after a rejected
membership measurement could avoid executing the remaining transport gates.
This capability is supported by platforms such as Quantinuum
Helios~\cite{Ransford2026Helios}, but was not available in the present
hardware experiment. For the same projected circuit in the noiseless case,
early termination leaves the accepted proposal distribution and hence the
final CI energy at a given accepted-sample count unchanged. Comparing the
final selected-space performance of projected and unprojected proposals is
beyond the present demonstration and remains a direction for ansatz refinement.

\subsection*{S1.E. Candidate ranking and selection}

Ordinary EN scores measure each candidate's direct coupling to the baseline.
CEN2 also includes contributions transmitted through other candidates,
allowing determinants with zero direct residual to receive a nonzero score.
Within each of the HCI and HCI+QRE comparisons, changing the ranking from
EN to CEN2 leaves the candidate pool, baseline coefficients, final dimension,
spin-completion rule, and eigensolver unchanged.
Candidates are selected as complete spin bundles at fixed spatial
occupation and $M_S=0$.

For a candidate determinant $\ket{D_I}$, define
\begin{equation}
 r_I=\sum_{J\in\mathcal V_0}H_{IJ}C_J^{(0)},\qquad
 \Delta_I=E_0-H_{II}.
 \label{eq:supp-direct}
\end{equation}
For both ordinary EN and CEN2, we impose a lower bound $\delta=10^{-12}$~Ha
on the denominator magnitude while preserving its sign to avoid division
by very small energy differences. Thus $H_{II}-E_0$ is replaced by
\begin{equation}
 d_I^{\mathrm{reg}}=
 \operatorname{sgn}(H_{II}-E_0)
 \max\{|H_{II}-E_0|,\delta\},
 \label{eq:supp-denominator}
\end{equation}
with the positive-sign convention at an exactly zero raw denominator, and the
signed amplitude denominator is $\Delta_I^{\mathrm{reg}}=-d_I^{\mathrm{reg}}$.
The sign is retained in every propagated amplitude.  Absolute denominators
are used only in the final EN-like scores.

CEN2 starts from the ordinary EN amplitudes
$a_I^{(0)}=r_I/\Delta_I^{\mathrm{reg}}$ and updates them twice using
couplings between candidates in the fixed pool $\mathcal P$. The first
update collects contributions from directly coupled candidates. The second
repeats this propagation using the updated amplitudes. Throughout, the
baseline energy and coefficients remain fixed. Only off-diagonal Hamiltonian
elements enter these updates, because the diagonal elements are already
included in the denominators:
\begin{align}
 a_I^{(0)}&=\frac{r_I}{\Delta_I^{\mathrm{reg}}},\\
 b_I^{(1)}&=\sum_{J\in\mathcal P\setminus\{I\}}H_{IJ}a_{J}^{(0)},&
 a_I^{(1)}&=\frac{r_I+b_I^{(1)}}{\Delta_I^{\mathrm{reg}}},\\
 b_I^{(2)}&=\sum_{J\in\mathcal P\setminus\{I\}}H_{IJ}a_{J}^{(1)},&
 \widetilde r_I^{(2)}&=r_I+b_I^{(2)}.
 \label{eq:supp-cen2}
\end{align}
CEN2 ranks determinants by
\begin{equation}
 S_I^{\mathrm{CEN2}}=
 \frac{|\widetilde r_I^{(2)}|^2}{|\Delta_I^{\mathrm{reg}}|}.
 \label{eq:supp-cen2-score}
\end{equation}
The direct residual $r_I$ and the propagated contribution $b_I^{(2)}$ are
added before squaring, allowing their signs to produce enhancement or
cancellation.
The two updates reach paths with at most two candidate--candidate couplings
beyond the initial baseline--candidate coupling, giving up to three
Hamiltonian couplings from the baseline. A candidate without such a path has zero
individual score, although a positive-score spin-bundle partner can admit it
to the final space.

Both propagation steps use the same sparse set of Hamiltonian couplings
between candidates, so their computational cost scales linearly with the
number of nonzero couplings.

We sum the determinant scores within each spin bundle and select complete
bundles to supply $F(R)=20{,}000-B(R)$ additions. Bundles are ranked by their
summed scores, with ties resolved by a fixed ordering of the configurations.

For the diagnostic in Fig.~\ref{fig:h12}(b), a selected source-exclusive
determinant is classified as zero-direct-residual when its sparse baseline
residual $r_I$ is identically zero.

\subsection*{S1.F. Final variational diagonalization}

We diagonalize the Hamiltonian once in the space containing the fixed
baseline and selected additions and obtain the lowest singlet state.
The resulting energies are variational values for matched
20,000-determinant spaces and are used in the main-text figures.

Fig.~\ref{fig:h12}(a) compares the proposal sources and ranking methods.
The HCI+ordinary-EN curve is omitted because it nearly coincides with
HCI+CEN2: the largest energy lowering from CEN2 reranking of the HCI pool
is 0.44~mHa at 1.7~\AA. For the HCI+QRE pool, this lowering reaches
10.80~mHa at $R=1.8$~\AA. QRE+CEN2 remains above the FCI energy at every
geometry, as required by the variational principle.

\begin{figure}[htbp]
 \centering
 \includegraphics[width=0.65\textwidth]{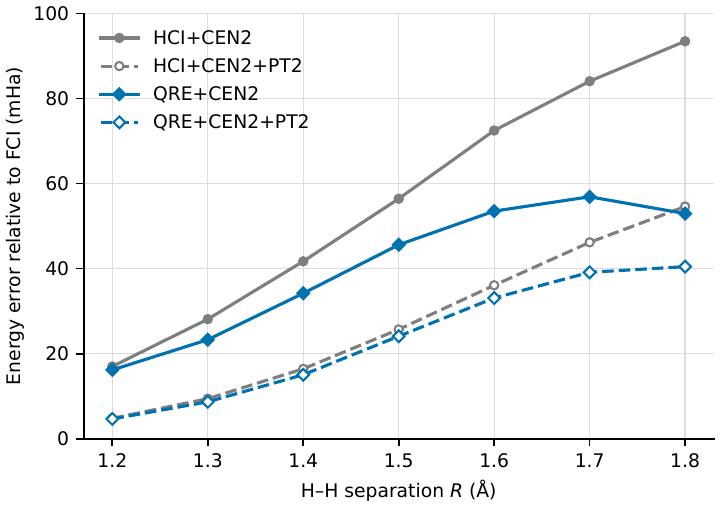}
 \caption{Variational and EN-PT2-corrected energy errors relative to FCI
 for HCI+CEN2 and QRE+CEN2 in H$_{12}$/STO-3G. All selected spaces contain
 $N=20{,}000$ determinants. Solid curves show variational errors; dashed
 curves show errors after EN-PT2 correction of the same final states.
 Corrected energies are not variational bounds.}
 \label{fig:supp-pt2}
\end{figure}

We next assess how EN-PT2 changes the comparison between the final
HCI+CEN2 and QRE+CEN2 states, using the correction procedure in End Matter.
CEN2 determines candidate selection, while EN-PT2 corrects the final energy
without changing the selected space. As shown in Fig.~\ref{fig:supp-pt2},
the QRE energy advantage remains at all seven geometries after correction.
At $R=1.8$~\AA, the HCI+CEN2 and QRE+CEN2 errors decrease from 93.45 and
52.95~mHa to 54.59 and 40.47~mHa, respectively, leaving a 14.12~mHa
advantage for QRE.

\section*{S2. Multi-generation HCI proposal-search diagnostic}

We test whether further classical determinant discovery closes the energy
gap to QRE proposals by extending ordinary HCI search through five
generations from the same protected baseline. For this discovery step,
HCI grows an auxiliary CI space and updates its wavefunction after each
expansion. The original baseline, coefficients, energy, and QRE statevector
pool remain fixed for the final CEN2 comparisons.

At each generation, we evaluate the HCI score in Eq.~\eqref{eq:hci} using
the current wavefunction in the auxiliary space and construct a
30,000-determinant candidate pool using the rules of Sec.~S1.C. Up to 25
eligible spin bundles are added to the auxiliary space, which is then
rediagonalized. This number was fixed by a pilot run without tuning against
FCI outcomes. The accumulated proposal pool retains all distinct external
candidates from these pools, including those not added to the auxiliary
space, and can therefore exceed 30,000 determinants. The first generation
reproduces the one-shot HCI pool.

At each depth $K=1,\ldots,5$, the fixed QRE pool, cumulative HCI pool,
and their union are ranked independently by CEN2 using the original
$\mathcal V_0,C_J^{(0)},E_0$. Each selected space is diagonalized once at a
common final dimension: $N=20{,}000$, except at 1.2~\AA, where the QRE-only
pool limits the comparison to $N=16{,}396$. The QRE-only branch here is
distinct from the primary QRE result in Fig.~\ref{fig:h12}, which uses the
HCI+QRE union at $N=20{,}000$ throughout.

\begin{figure}[t]
 \centering
 \includegraphics[width=0.48\textwidth]{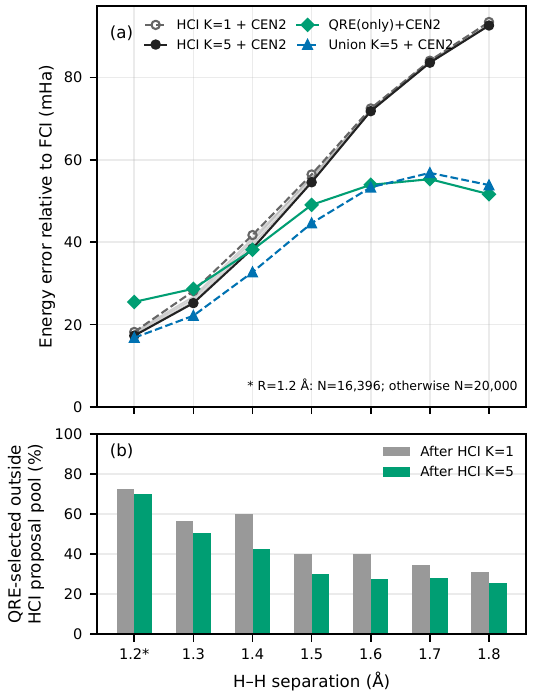}
 \caption{Multi-generation HCI proposal-search diagnostic for the same
 threshold-converged H$_{12}$ baselines and fixed QRE statevector pools.
 (a) Final variational error relative to FCI for QRE(only)+CEN2, HCI-only at
 generations $K=1$ and $K=5$, and the $K=5$ union, each ranked once by CEN2
 against the original baseline. Faint lines show HCI-only at $K=2$--4.
 (b) Percentage of QRE-only-branch selected additions absent from the
 cumulative capped HCI candidate pool after $K=1$ or $K=5$; the denominator
 is the geometry-specific number of QRE additions $N-B(R)$. The HCI pool
 includes candidates not added to the auxiliary space. The common dimension is $N=16{,}396$ at 1.2~\AA\ and $N=20{,}000$ elsewhere.}
 \label{fig:supp-multigen}
\end{figure}

As shown in Fig.~\ref{fig:supp-multigen}, HCI-only proposals match or improve on
QRE-only proposals at $K=1$ for 1.2--1.3~\AA. They do not match QRE-only within
five generations for 1.4--1.8~\AA; the $K=5$ energy gap is 0.30~mHa at 1.4~\AA\
and 40.97~mHa at 1.8~\AA. Some additions selected from QRE remain absent from
the cumulative HCI pool after five generations. The union can give a higher
energy than QRE-only because CEN2 may replace candidates when ranking the larger
pool at fixed final dimension.

Thus, extending HCI discovery through five generations does not eliminate
the energy advantage of QRE-only proposals at stretched geometries, whereas
HCI is already competitive at short separations. QRE therefore supplies
useful complementary candidates under the tested policy of adding up to
25 spin bundles per generation.

\end{document}